\pdfoutput=1
\documentclass{article}
\usepackage{spconf,amsmath,graphicx}
\usepackage{cite,xcolor,textcomp,algorithmic,epsfig,multirow,hhline}
\usepackage{amssymb,amsfonts}

\title{An Improved Uncertainty Propagation Method for Robust I-vector Based Speaker Recognition}
\twoauthors
{Dayana Ribas}
{ViVoLab, Aragon Institute for \\
    Engineering Research (I3A), Spain}
{Emmanuel Vincent}
{Universit\'e de Lorraine, CNRS, \\
    Inria, LORIA, F-54000 Nancy, France}

\begin{document}
\ninept

\maketitle
\begin{abstract}
The performance of automatic speaker recognition systems degrades when facing distorted speech data containing additive noise and/or reverberation. Statistical uncertainty propagation has been introduced as a promising paradigm to address this challenge. So far, different uncertainty propagation methods have been proposed to compensate noise and reverberation in i-vectors in the context of speaker recognition. They have achieved promising results on small datasets such as YOHO and Wall Street Journal, but little or no improvement on the larger, highly variable NIST Speaker Recognition Evaluation (SRE) corpus. In this paper, we propose a complete uncertainty propagation method, whereby we model the effect of uncertainty both in the computation of unbiased Baum-Welch statistics and in the derivation of the posterior expectation of the i-vector. We conduct experiments on the NIST-SRE corpus mixed with real domestic noise and reverberation from the CHiME-2 corpus and preprocessed by multichannel speech enhancement. The proposed method improves the equal error rate (EER) by 4\% relative compared to a conventional i-vector based speaker verification baseline. This is to be compared with previous methods which degrade performance.
\end{abstract}


\vspace{3mm}
\section{Introduction}
Uncertainty propagation has emerged as a paradigm for robust signal processing whereby the data are not treated as point estimates anymore, but as a parametric posterior distribution, typically approximated as a Gaussian. This approach provides a principled framework to deal with the loss of information due to signal distortion -- epistemic uncertainty -- or to the finite number of data points -- aleatoric uncertainty --. The uncertainty is represented by a set of scalar variances or covariance matrices, which are  first estimated on the data, and then propagated through the subsequent processing steps in order to compensate for the effect of uncertainty on the computed quantities and ultimately improve the system performance \cite{Arrowood2002,Becerra2002,Deng2005,Liao2005}.


In speaker recognition, the uncertainty propagation approach has gained traction motivated by the uncertain nature of the system pipeline when the application scenario tends to more real-world situations. The necessity of improving the system robustness in noisy environments has inspired the development of epistemic uncertainty approaches for speaker modeling based on Gaussian mixture models \cite{Zhao2012,Ozerov2013} and, more recently, i-vectors \cite{Yu2014,Ribas2015a}. Other work based on the aleatoric uncertainty concept has also given rise to uncertainty propagation approaches for speaker recognition. However, these approaches focused on the issue of computing representations with insufficient data, caused by utterances with different, possibly short durations \cite{Hautamaki2013,Stafylakis2013,Kenny2013,Cumani2014,Saeidi2015,Kenny2016,Mak2017}.  

This paper pursues the same line as our preliminary study \cite{Ribas2015a}, that considered an epistemic uncertainty propagation approach for noise-robust text-independent speaker verification using a system based on i-vectors \cite{Dehak2011} and probabilistic linear discriminant analysis (PLDA) \cite{Garcia2011}. Despite the recent introduction of deep learning based modules in the speaker recognition pipeline, reports of the last NIST Speaker Recognition Evaluation (SRE) in 2016 \cite{I4U2016} and the experience in the recent campaign NIST-SRE 2018\footnote{https://www.nist.gov/itl/iad/mig/nist-2018-speaker-recognition-evaluation} evidenced that the i-vector-PLDA approach still performs among the best systems of the state-of-the-art. It is also the focus of current challenges in the field such as the 2018 Multi-target Speaker Detection and Identification Challenge Evaluation\footnote{http://mce.csail.mit.edu/}.        

In \cite{Ribas2015a}, we proposed a method to estimate and propagate the residual uncertainty after multichannel speech enhancement from the acoustic features to the i-vectors. Specifically, we modified the posterior probability of each Gaussian mixture component to obtain unbiased Baum-Welch (BW) statistics. Preliminary experiments yielded good results on the Wall Street Journal (WSJ) corpus, but little or no improvement on a subset of the NIST-SRE 2008 corpus, similarly to the findings of Yu et al.\ \cite{Yu2014} on the YOHO and NIST-SRE 2010 datasets. We studied the causes of this under-performance on NIST-SRE and found out that the high variability of NIST-SRE makes it intrinsically harder to obtain accurate BW statistics.

In this paper, we propose a new uncertainty propagation method that models the effect of epistemic uncertainty both in the computation of the BW statistics and in the derivation of the i-vector. This method provides a more complete strategy towards compensating for background noise in all steps of the i-vector computation process. Furthermore, this study contributes to clarifying how the i-vector speaker representations are affected by the environmental distortion. The results are evaluated on a subset of the NIST-SRE 2008 corpus mixed with real domestic noise and reverberation from the CHiME-2 corpus \cite{Chime2013}.

Section \ref{sec:ivector} recalls the i-vector computation process. Section \ref{sec:uncert} provides a novel analysis of the limitations of related previous works and Section \ref{sec:proposal} introduces the proposed method. Speaker verification experiments are conducted in Section \ref{sec:expe}. The results are reported and discussed in Section \ref{sec:disc}. Finally, the conclusions of the study and future work are presented in Section \ref{sec:concl}.

\section{I-vector computation}
\label{sec:ivector}
Front-end factor analysis \cite{Dehak2011} relies on a universal background model (UBM) that is a mixture of $C$ Gaussian components indexed by $c$. Denoting by $F$ the feature dimension, the $CF\times 1$ supervector $M(u)$ for one utterance $u$ is expressed as 
\begin{equation}
M(u) = m + Tw(u) + \epsilon(u)
\label{eqModel}
\end{equation}
where $m$ consists of the means $m_c$ of all UBM components, $T$ is the $CF\times D$ low-rank total variability matrix, $w(u)$ is the $D\times 1$ vector of total factors or \emph{i-vector}, and $\epsilon(u)$ represents the residual data variability not captured by $T$. The i-vector is modeled as a zero-mean standard Gaussian random vector. It is obtained by computing the posterior expectation of $w(u)$ over the feature sequence $\{y_1,\dots,y_L\}$, with $L$ the number of time frames:
\begin{equation}
\mathbb{E}[w(u)] = (I + T'V^{-1}N(u)T)^{-1} T'V^{-1}\hat{F}(u).
\label{eqIvector}
\end{equation}
In this equation, $N(u)$ is a $CF\times CF$ diagonal matrix with diagonal blocks $N_c(u)I$ where $N_c(u)$ are the zeroth-order BW statistics for all components $c$, $\hat{F}(u)$ is a $CF\times 1$ supervector obtained by concatenating the centralized first-order BW statistics $\hat{F}_c(u)$, $V$ is the diagonal $CF\times CF$ covariance matrix of $\epsilon(u)$, and $'$ denotes matrix transposition. The BW statistics are given by
\begin{align}
N_c(u) &= \sum_{t=1}^L\gamma_{t}(c)\label{eqN}\\
\hat{F}_c(u) &= \sum_{t=1}^L\gamma_{t}(c) (y_{t}-m_{c})\label{eqF}
\end{align}
where 
\begin{equation}
\gamma_{t}(c) = \frac{\pi_{c}\,\mathcal{N}(y_{t}|\mu_{c},\Sigma_{c})}{\sum^{C}_{i=1}\pi_{i}\,\mathcal{N}(y_{t}|\mu_{i},\Sigma_{i})}
\label{eqGamma2}
\end{equation}\\
is the posterior probability of the $c$-th UBM component, as obtained from its mean $m_c$, covariance $\Sigma_c$ and weight $\pi_c$.

Note that \eqref{eqIvector} can be equivalently rewritten in terms of the ``normalized'' statistics $\tilde{N}(u)$ and $\tilde{F}(u)$ as
\begin{equation}
\mathbb{E}[w(u)] = (I + T'\tilde{N}(u)T)^{-1} T'\tilde{F}(u).
\label{eqIvector2}
\end{equation}
where $\tilde{N}(u)=V^{-1}N(u)$ is a diagonal matrix with diagonal blocks $\tilde{N}_c(u)=N_c(u)V_c^{-1}$, $\tilde{F}(u)=V^{-1}\hat{F}(u)$ is obtained by concatenating $\tilde{F}_c(u)=V_c^{-1}\hat{F}_c(u)$, and $V_c$ is the $c$-th diagonal block of $V$. The multiplication by $V_c^{-1}$ can be distributed at each time $t$ of the summations in \eqref{eqN} and \eqref{eqF}. The purpose of this rewriting will become clear in the next section.

\section{Uncertainty propagation to the i-vector}
\label{sec:uncert}
Let us assume that we now observe a corrupted speech signal $x_t$ involving noise and/or reverberation. Using a speech enhancement algorithm together with an uncertainty estimation technique, the posterior probability of the clean speech features $y_t$ can be modeled as \cite{Kolossa2011}
\begin{equation}
p(y_t|x_t)=\mathcal{N}(y_t|\bar{y}_{t},\bar{\Sigma}_{t})
\end{equation}
with $\bar{y}_{t}$ the enhanced features and $\bar{\Sigma}_{t}$ the uncertainty covariance matrix at time $t$. In other words, $\bar{\Sigma}_{t}$ is the covariance of the estimation error between the enhanced features and the (unknown) clean features at a given time.

To the best of our knowledge, the studies in \cite{Yu2014,Ribas2015a} are the only ones exploiting this model for noise and reverberation robustness in i-vector based speaker recognition systems, while other works focused on earlier, now deprecated systems. They proposed two different ways to propagate the uncertainty from the enhanced features to the i-vectors.

\subsection{Uncertainty propagation through the front-end factor analysis model}
The authors in \cite{Yu2014} considered the generative data model corresponding to \eqref{eqModel}. By integrating over the unknown clean features, they accounted for the impact of uncertainty on the expression of the joint posterior probability. They derived the posterior expectation of $w(u)$ over the feature sequence in a similar way to \eqref{eqIvector2} as\footnote{For the sake of clarity, we modified the notations in \cite{Yu2014} for consistency with the above.}
\begin{equation}
\mathbb{E}[w_\mathrm{unc}(u)] = (I + T'\tilde{N}_\mathrm{unc}(u)T)^{-1} T'\tilde{F}_\mathrm{unc}(u).
\label{eqIvectorUNCYu}
\end{equation}
$\tilde{N}_\mathrm{unc}(u)$ becomes the $CF\times CF$ block-diagonal matrix with diagonal blocks
\begin{equation}
\tilde{N}_{\mathrm{unc},c}(u)=\sum_{t=1}^L\gamma_{t}(c)V_{\mathrm{unc},c,t}^{-1},\label{eqNUNCYu}
\end{equation}
$\tilde{F}_\mathrm{unc}(u)$ is the $CF\times 1$ supervector obtained by concatenating
\begin{equation}
\tilde{F}_{\mathrm{unc},c}(u)=\sum_{t=1}^L\gamma_{t}(c)V_{\mathrm{unc},c,t}^{-1}(\bar{y}_{t}-m_{c}),\label{eqFUNCYu}
\end{equation}
and $V_{\mathrm{unc},c,t}$ is the total covariance of the residual variability and the uncertainty:
\begin{equation}
V_{\mathrm{unc},c,t}=V_c+\bar{\Sigma}_{t}.\label{eqVunc}
\end{equation}

In order to compute the posterior probability of the $c$-th UBM component, the authors substituted the clean features ${y}_{t}$ in \eqref{eqGamma2} by the enhanced features $\bar{y}_{t}$: 
\begin{equation}
\gamma_{t}(c) = \frac{\pi_{c}\,\mathcal{N}(\bar{y}_{t}|\mu_{c},\Sigma_{c})}{\sum^{C}_{i=1}\pi_{i}\,\mathcal{N}(\bar{y}_{t}|\mu_{i},\Sigma_{i})}\label{eqGammaYu}
\end{equation}
This expression does not account for the difference between the enhanced features and the clean features, hence the BW statistics are biased. Indeed, it is well known from the field of speech recognition that using enhanced data as inputs to an acoustic model trained on clean data often results in poor recognition performance due to the residual distortions in the enhanced data \cite{Deng2005,Kolossa2011}. We conclude that the uncertainty is not fully propagated to the i-vector domain, and therefore the obtained i-vectors remain affected by the data distortion. This may be the reason for the poor results achieved by this method on the NIST-SRE 2010 corpus, even in the situation when \emph{oracle} (ideal) uncertainty estimates are used \cite{Yu2014}.

\subsection{Uncertainty propagation through the UBM}
The study in \cite{Ribas2015a} presented an algorithm to compute unbiased BW statistics instead. By considering the generative data model associated with the UBM and integrating over the unknown clean features, the likelihood $p(y_t|c)=\mathcal{N}(y_{t}|\mu_{c},\Sigma_{c})$ of the $c$-th UBM component is classically substituted by \cite{Deng2005}
\begin{equation}
p(x_t|c)= \mathcal{N}(\bar{y}_{t}|\mu_{c},\Sigma_{\mathrm{unc},c,t})
\label{eqloglikelihood}
\end{equation}
where
\begin{equation}
\Sigma_{\mathrm{unc},c,t} = \Sigma_{c} + \bar{\Sigma}_{t}.
\end{equation}

The authors considered the i-vector computation in \eqref{eqIvector} as a deterministic operation and consequently derived the i-vector as
\begin{equation}
\mathbb{E}[w_\mathrm{unc}(u)] = (I + T'V^{-1}N_\mathrm{unc}(u)T)^{-1} T'V^{-1}\hat{F}_\mathrm{unc}(u)
\label{eqIvectorUNCRibas}
\end{equation}
where $N_\mathrm{unc}(u)$ is a diagonal matrix with blocks $N_{\mathrm{unc},c}(u)I$, $\hat{F}_\mathrm{unc}(u)$ is the concatenation of $\hat{F}_{\mathrm{unc},c}(u)$, and the unbiased BW statistics \cite{Ozerov2013} are given by
\begin{align}
\gamma_{\mathrm{unc},t}(c) &= \frac{\pi_{c}\,\mathcal{N}(\bar{y}_{t}|\mu_{c},\Sigma_{\mathrm{unc},c,t})}{\sum^{C}_{i=1}\pi_{i}\,\mathcal{N}(\bar{y}_{t}|\mu_{i},\Sigma_{\mathrm{unc},i,t})}\label{eqGammaUNC}\\
N_\mathrm{unc,\emph{c}}(u) &= \sum_{t=1}^L\gamma_{\mathrm{unc},t}(c)\label{eqNunc}\\
\hat{F}_\mathrm{unc,\emph{c}}(u) &= \sum_{t=1}^L\gamma_{\mathrm{unc},t}(c) W_{c,t} (\bar{y}_{t}-m_{c})\label{eqFunc}
\end{align}
with $W_{c,t}$ the Wiener filter \cite{Bishop2006} defined as
\begin{equation}
W_{c,t} = \Sigma_{c}\Sigma_{\mathrm{unc},c,t}^{-1}.\label{eqW}
\end{equation}

Despite the use of unbiased BW statistics, the resulting i-vectors are not fully compensated for the uncertainty at the input. Since the i-vector computation is not according to the modifications introduced to the speaker model due to the propagation of the uncertainty, it raises a mismatch that negatively impacts the system performance. Indeed, i-vector computation is not a deterministic operation and it must be modified to account for this uncertainty. This may partly explain the poor results reported on the NIST-SRE 2008 corpus in \cite{Ribas2015a}, in addition to the high variability of this corpus which makes it harder to obtain meaningful BW statistics (see study in \cite{Ribas2015a}).

\section{Proposed method}
\label{sec:proposal}
The analysis of previous works on uncertainty propagation for i-vector based speaker recognition has uncovered the fact that the uncertainty is propagated either through the front-end factor analysis model or through the UBM. In order to fully compensate i-vectors for the distortion of the input data, we propose to propagate the uncertainty through both models. I-vectors are then computed in the same way as in \eqref{eqIvectorUNCYu}, albeit with unbiased normalized statistics. Specifically, $\tilde{N}_\mathrm{unc}(u)$ is now the block-diagonal matrix with diagonal blocks
\begin{equation}
\tilde{N}_{\mathrm{unc},c}(u)=\sum_{t=1}^L\gamma_{\mathrm{unc},t}(c)V_{\mathrm{unc},c,t}^{-1}\label{eqNUNCProp}
\end{equation}
and $\tilde{F}_\mathrm{unc}(u)$ is obtained by concatenating
\begin{equation}
\tilde{F}_{\mathrm{unc},c}(u)=\sum_{t=1}^L\gamma_{\mathrm{unc},t}(c)V_{\mathrm{unc},c,t}^{-1}W_{c,t}(\bar{y}_{t}-m_{c}),\label{eqFUNCProp}
\end{equation}
where $\gamma_{\mathrm{unc},t}(c)$, $V_{\mathrm{unc},c,t}$, and $W_{c,t}$ are defined in \eqref{eqGammaUNC}, \eqref{eqVunc}, and \eqref{eqW}, respectively.\\

This time, the enhanced data $\bar{y}_{t}$ and the uncertainty $\bar{\Sigma}_{t}$ are both used to compute the unbiased BW statistics. Furthermore, the i-vector obtained from the computation of the posterior expectation of $w(u)$ also considers the influence of $\bar{\Sigma}_{t}$.\\

\begin{figure}[h!]
\centerline{\epsfig{figure=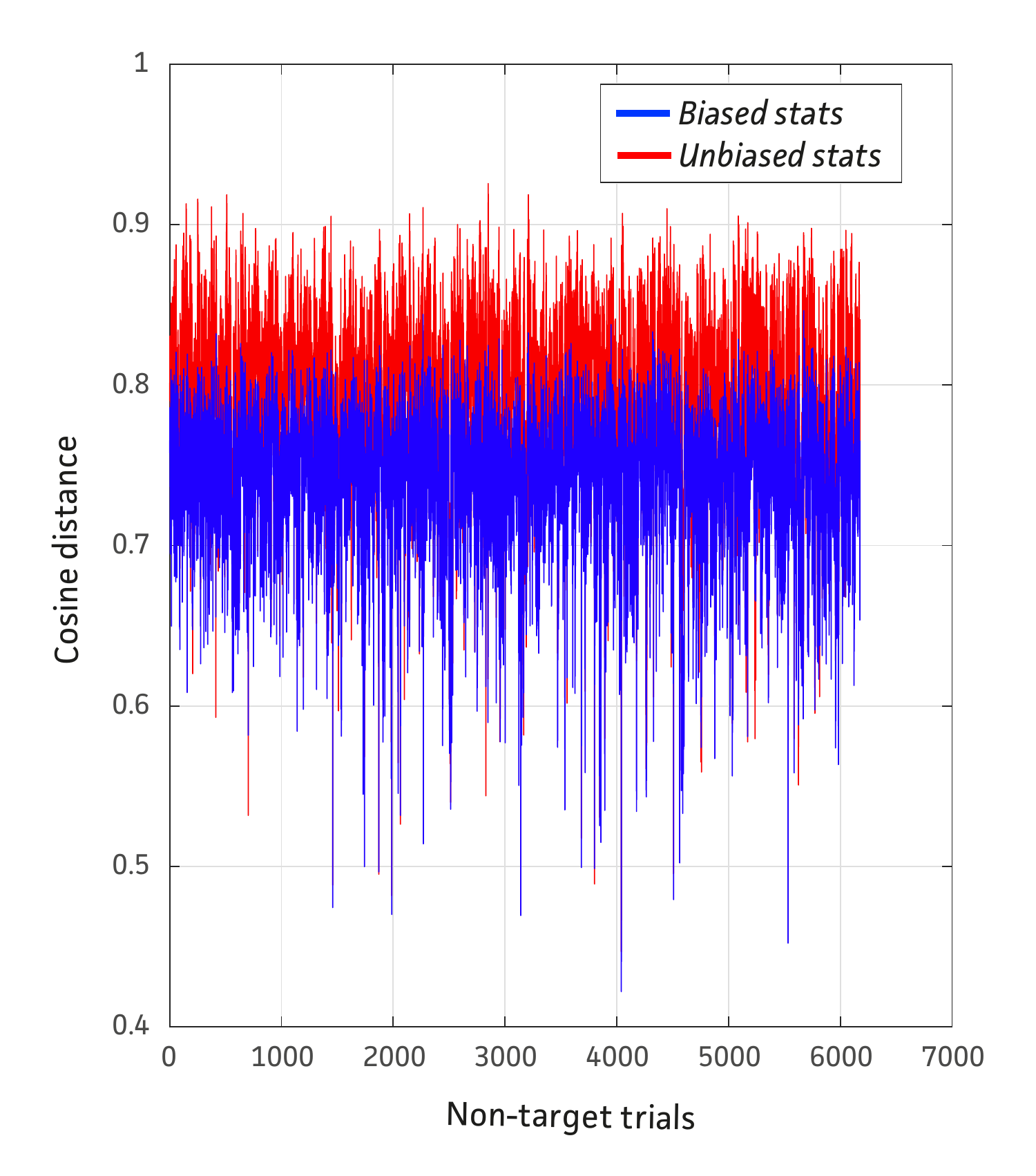,width=80mm}}
\caption{{\it Cosine distance for the unbiased and biased $\hat{F}$ statistics in non-target trials.}}
\label{figFcosine}
\end{figure}

In order to assess the impact of unbiased BW statistics independently of the i-vector computation process, we compared them to biased BW statistics as follows. We chose the unnormalized $\hat{F}$ statistics as a rough representation of the speaker characteristics before i-vector computation. We computed the cosine distance between the $\hat{F}$ vectors of each trial of the dataset used in the experimental setup (see section \ref{dataset}). The results displayed in Fig.\ \ref{figFcosine} show higher cosine distances for the unbiased statistics in non-target trials. The average cosine distance over all non-target trials is $0.7939$ for the unbiased $\hat{F}$ statistics vs.\ $0.7422$ for the biased $\hat{F}$ statistics. This suggests that non-target utterances can be better separated using unbiased statistics. The poor results reported on the NIST-SRE 2008 corpus in \cite{Ribas2015a} using i-vector based cosine distances suggest that the benefit of unbiased BW statistics observed when considering the statistics themselves was lost when considering i-vectors instead. Overall, this suggests that unbiased BW statistics and uncertainty-aware i-vector computation are both necessary to achieve good results.

\section{Experimental setup}
\label{sec:expe}
\subsection{Dataset description}
\label{dataset}
We compare the proposed method with \cite{Yu2014} and \cite{Ribas2015a} using the same experimental setup as in \cite{Ribas2015a}. The speech signals are male conversations in English from the NIST-SRE 2004 and 2005 corpora for the training stage (3285 speech signals from 262 speakers), and from the NIST-SRE 2008 \emph{short2} and \emph{short3} corpora for the evaluation stage (enrollment: 470 utterances, test: 671 utterances). A total of 6615 verifications were performed on the \emph{det7} condition of NIST-SRE. 

We used two-channel noise and reverberation from Track 1 of the 2nd CHiME Challenge \cite{Chime2013}, recorded in a real domestic environment. Each training or evaluation signal was convolved with one of 121 two-microphone room impulse responses with a reverberation time of 0.3 seconds. Moreover, the enrollment and test signals were mixed with a random segment of real background noise including, e.g., voices, TV, game console, cutlery sounds, and footsteps. This mixing process resulted in noisy speech signals with different SNRs ranging from about -10 to +20 dB with an average of 6.1 dB. 

\subsection{Speech enhancement and uncertainty estimation}
For speech enhancement, we applied the CHiME-2 recipe \cite{Fasst2014} of the Flexible Audio Source Separation Toolbox (FASST) \cite{Ozerov2012}. This recipe was designed to reduce noise, but not reverberation. Hence we consider $y_t$ to be the reverberated noiseless data. For uncertainty estimation, in order to assess the potential of each method, we used the oracle diagonal uncertainty covariance matrices $\bar{\Sigma}_{t}=\mathrm{diag}(y_t-\bar{y}_t)^2$ \cite{Kolossa2011}. However, for a real application there are several uncertainty estimators \cite{Tran2015}. 

\subsection{Speaker recognition system}
The features consist of 19 Mel frequency cepstral coefficients (MFCCs), the log-energy, and their first- and second-order derivatives, followed by voice activity detection (VAD) and cepstral mean and variance normalization (CMVN) \cite{Handbook2008}. The UBM ($C=512$) and the $T$ matrix ($D=400$) were trained on either clean or reverberated noiseless data \cite{Ribas2015}. Since the chosen speech enhancement method reduces noise but not reverberation, training on reverberated speech provides better results than training on clean speech. The i-vectors are centered, whitened and length-normalized, and subsequently projected with 330-dimensional LDA. Classification relies on Gaussian PLDA \cite{Garcia2011}. 

\section{Results and discussion}
\label{sec:disc}
The speaker verification results expressed in terms of equal error rate (EER) are presented in Table \ref{tableEER}. For comparison purposes, the first two rows show the results obtained when training and testing on clean (original NIST) signals or when training and testing on reverberated noiseless signals. The following rows show the results obtained when training on reverberated noiseless signals and testing on noisy or enhanced signals. The last three rows correspond to the two previous uncertainty propagation methods reviewed in Section \ref{sec:uncert} and to the proposed one in Section \ref{sec:proposal}. Yu et al.'s method \cite{Yu2014} was tested based on an implementation from the authors.

\begin{table} [th]
\caption{\label{tableEER} {\it Speaker recognition results on NIST-SRE.}}
\vspace{1mm}
\centerline{
\begin{tabular}{|c|c|c|} 
\hline
\textbf{Training set} & \textbf{Evaluation set} & \textbf{EER (\%)} \\ \hhline{===} 
Clean & Clean & 3.19 \\ \hline 
Reverberated & Reverberated & 4.33 \\ \hhline{===} 
Reverberated & Noisy & 31.85 \\ \hline 
Reverberated & Enhanced & 10.48 \\ \hline 
Reverberated & Uncertainty propagation \cite{Yu2014} & 11.39 \\ \hline
Reverberated & Uncertainty propagation \cite{Ribas2015a} & 10.69 \\ \hline 
Reverberated & Proposed uncertainty propagation & \textbf{10.02} \\ \hline
\end{tabular}}
\end{table}

The best achievable EER on reverberated data (without performing dereverberation) is 4.33\%. Background noise strongly degrades it up to 31.85\%. Multichannel speech enhancement improves it down to 10.48\%. Previous uncertainty propagation methods do not outperform this baseline; actually, the method in \cite{Yu2014} significantly degrades it. Only the proposed method manages to improve performance down to an EER of 10.02\%. This corresponds to a 4\% relative improvement with respect to the speech enhancement baseline without uncertainty propagation. Although this might seem limited, it is to the best of our knowledge the first time an improvement in terms of noise robustness is reported for i-vector based systems by means of uncertainty propagation on a large NIST-SRE corpus.

\begin{figure}[h!]
\centerline{\epsfig{figure=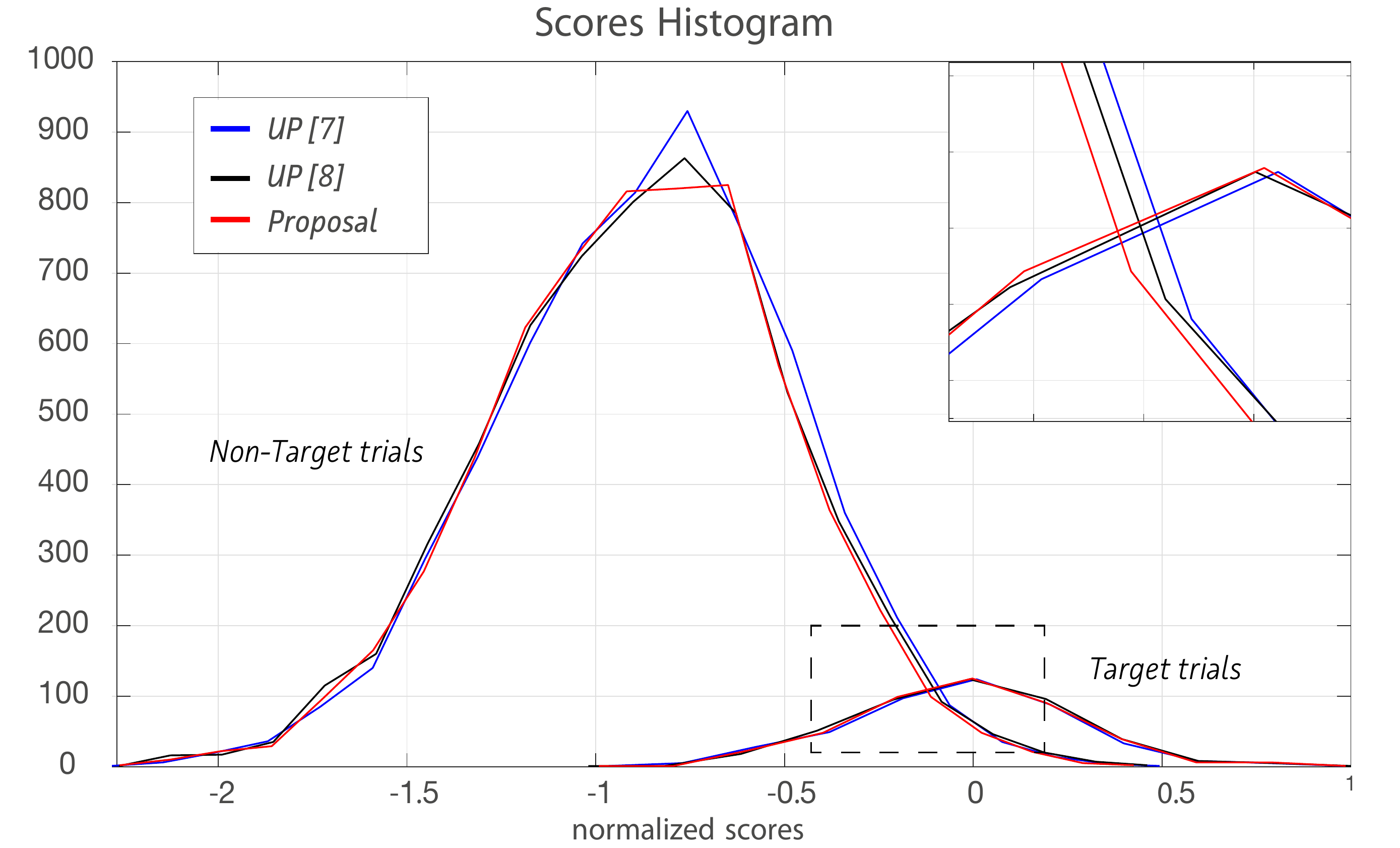,width=90mm}}
\caption{{\it Histogram of scores for non-target (left) and target trials (right). The top right corner is a zoomed view of the dashed area.}}
\label{figscores}
\end{figure}

Figure \ref{figscores} shows the histograms of scores for the three uncertainty propagation algorithms in Table \ref{tableEER}. The scores are normalized using the maximum score value across the three sets. The scores for non-target trials obtained with the proposed method appear to be smallest among the three tested systems. Indeed the corresponding curve is the leftmost one in the zoomed view. This behavior is consistent with our findings regarding unbiased vs.\ biased statistics in Section \ref{sec:proposal}. This indicates there is a wider separation between non-target and target trials in the system using the proposed uncertainty propagation algorithm, hence the performance improvement obtained in Table \ref{tableEER}. 

\section{Conclusions}
\label{sec:concl}
In this paper, we proposed a new method for the propagation of uncertainty due to speech distortion from the input data to the i-vectors in a speaker recognition framework. We analyzed the limitations of previously proposed methods, and introduced a more complete propagation method that operates both on the UBM and on the front-end factor analysis model. We evaluated the speaker verification performance on the NIST-SRE 2008 corpus mixed with real domestic noise and reverberation and obtained a 4\% relative EER improvement compared to a speech enhancement baseline without uncertainty propagation, while previous methods degraded performance. Future work will target the extension of the uncertainty propagation approach to the full speaker verification pipeline including the PLDA and the joint handling of epistemic uncertainty due to signal distortion and aleatoric uncertainty due to finite signal size. Also we plan to extend the experimental setup for including female speakers. 

\section{Acknowledgments}
Authors would like to thank Dr. Chengzhu Yu for sharing the implementation of the UP referenced approach in \cite{Yu2014}. We are also grateful for the useful discussions with Dr. Antonio Miguel and Dr. Alfonso Ortega from ViVoLab. 
\bibliographystyle{IEEEbib}
\bibliography{biblio_short}

\end{document}